\documentclass[nohyper,12pt,a4paper]{JHEP}
\usepackage{epsfig}
\usepackage{psfrag}
\usepackage{graphicx}
\usepackage{amsfonts}

\def\beq{\begin{equation}}
\def\eeq{\end{equation}}
\def\bea{\begin{eqnarray}}
\def\eea{\end{eqnarray}}

\newcommand{\OO}{{\cal O}}
\newcommand{\rdot}{r^\prime}
\newcommand{\chip}{\chi^\prime}

\title{Thermodynamics of Large AdS Black Holes}

\author{Samuli Hemming\\
Academy of Finland\\
Vilhonvuorenkatu 6\\
P.O.Box 99, 00501 Helsinki, Finland\\
E-mail: \email{samuli@raunvis.hi.is}}

\author{L\'arus Thorlacius\\ 
University of Iceland, Science Institute \\
Dunhaga 3, 107 Reykjavik, Iceland\\
E-mail: \email{lth@hi.is}}

\abstract{
We  consider leading order quantum corrections to the geometry
of large AdS black holes in a spherical reduction of four-dimensional 
Einstein gravity with negative cosmological constant. The Hawking 
temperature grows without bound with increasing black hole mass, 
yet the semiclassical back-reaction on the geometry is relatively mild,
indicating that observers in free fall outside a large AdS black hole
never see thermal radiation at the Hawking temperature. 
The positive specific heat of large AdS black holes is a statement 
about the dual gauge theory rather than an observable property on 
the gravity side. Implications for string thermodynamics with an 
AdS infrared regulator are briefly discussed.
 }


\preprint{RH-05-2007}

\begin{document}




\section{Introduction}
Black holes play an important role in the
AdS/CFT correspondence 
\cite{Maldacena:1997re,Gubser:1998bc,Witten:1998qj}. 
In particular, the so-called large 
AdS--Schwarzschild black holes, whose horizon area is large
in units of the characteristic AdS length scale, correspond to 
high-temperature thermal states in the dual gauge 
theory \cite{Witten:1998zw,Hawking:1982dh}.
Another motivation for studying AdS black holes comes from 
string thermodynamics. The infrared instability of a self-gravitating
gas of strings can be regulated by embedding the system in an
asymptotically AdS background and in this context large
AdS--Schwarzschild black holes are the dominant
configurations at sufficiently high total 
energy \cite{Barbon:2001di}.

Large AdS--Schwarzschild black holes differ from ordinary
Schwarzschild black holes in asymptotically flat spacetime 
in a number of ways, but perhaps the most striking difference
is that the Hawking temperature of large AdS--Schwarzschild 
black holes grows as a function of the black hole mass
\cite{Hawking:1982dh}.
In this sense these black holes have positive specific heat.
Due to the curvature in the asymptotic AdS region, 
Hawking radiation from a black hole gets reflected back 
and at the quantum level one expects to find 
equilibrium configurations where a black hole emits Hawking 
radiation and absorbs particles from the environment at the
same rate. In Section~\ref{sec:semiclass} we consider semiclassical
corrections to the AdS--Schwarzschild black hole geometry,
which are obtained via spherical reduction of four-dimensional 
gravity. We look for static solutions to describe equilibrium
configurations involving large AdS--Schwarzschild black holes 
and study some of their physical properties.

Formally there is no upper limit on the mass of a large 
AdS--Schwarzschild black hole and the Hawking 
temperature can be arbitrarily high. Does this mean that
a large semiclassical black hole is in equilibrium with a 
high-temperature environment? Our answer is no, but 
before we get to that let us consider the question 
more carefully. First of all, in an asymptotically AdS geometry
there is infinite gravitational redshift as one goes to spatial
infinity and therefore the local temperature goes to zero 
asymptotically far from a AdS--Schwarzschild black hole, 
regardless of how high the Hawking temperature is.
The question is rather whether there is a
region near the black hole, {\it i.e.} within a proper distance 
of order the AdS length scale or so of the event horizon, 
where the local temperature becomes of order the 
Hawking temperature. 

The answer to this question 
depends on what kind of observer we have in mind. 
Let us first consider a fiducial observer, who is kept at rest
with respect to the static AdS--Schwarzschild coordinates.
Such an observer will indeed measure high temperatures
near a large AdS--Schwarzschild black hole but this
can be attributed to the acceleration required to keep
the observer from falling into the black hole and is the
same effect as found outside an ordinary Schwarzschild
black hole \cite{Unruh:1976db,Unruh:1982ic}. 
In particular, this {\it fiducial} temperature goes
to infinity as we approach the black hole event horizon
but we do not expect this to lead to strong back reaction
effects for AdS--Schwarzschild black holes anymore than
it does for Schwarzschild black holes.

It is more interesting to ask what temperature would be
recorded by an observer in free fall near the black hole.
This is the relevant temperature when we consider quantum 
corrections to the black hole geometry. If the black hole
is in equilibrium with a layer of high-temperature 
radiation, then the proper energy density is high in that
layer and one would expect large corrections to the geometry.
This would have to be reconciled with the fact 
that in the limit of large black hole mass 
the spacetime curvature in the horizon region does 
not grow beyond the AdS scale, as we will see
in section~\ref{sec:classical}, and thus {\it a priori\/} 
the size of quantum corrections is not expected
to grow with increasing black hole mass.

The question of the local temperature outside a
black hole also comes up in the context of string 
thermodynamics in an asymptotically AdS background. 
The Jeans instability undermines efforts to define a
thermodynamic limit in a gravitational theory such as
string theory but the troublesome infrared behavior 
can be regulated by introducing a negative cosmological 
constant. At weak coupling and low energies there exist 
spherically symmetric equilibrium configurations consisting 
of a gas of strings confined to a local region in an 
asymptotically AdS spacetime. 
The infrared regulated configurations are not spatially
homogeneous and therefore they do not describe a 
uniform thermal equilibrium state. They do, however, 
allow the notion of local thermal equilibrium
if the magnitude of the cosmological
constant is sufficiently small.

Since only massless modes of the string are excited 
at low energies the string gas consists to a very good 
approximation of massless radiation. Classical solutions 
of Einstein gravity with a negative cosmological constant
that describe spherically symmetric 'stars' made of 
self-gravitating radiation were found 
in \cite{Page:1985em,Hubeny:2006yu} and are briefly 
described in section~\ref{adsstars}. Analogous 
solutions can be found of the low-energy effective 
field theory in string theory. 
The Jeans instability resurfaces as a gravitational 
instability of the AdS star configurations at  energies of
order the AdS scale and at high energies 
the dominant configurations are large 
AdS--Schwarzschild black holes \cite{Barbon:2001di}.

At the semi-classical level, one still expects to find static 
solutions that describe stars made from self-gravitating 
radiation. It may be difficult to write them down in closed 
form but  their physical properties will be very similar to those 
of the corresponding classical solutions. One also expects
to find static solutions that describe large black holes in
equilibrium with Hawking radiation reflected from the
asymptotic AdS region and we look for such solutions in
section~\ref{sec:semiclass}. The transition from AdS stars to 
large black holes takes place at the AdS scale. 
Given that we want to have a description in terms
of (semi-)classical geometry we have to take the
AdS length large compared to the fundamental 
string length and this means that the AdS scale
transition temperature is far below the Hagedorn
temperature.\footnote{This is unfortunate 
in the sense that it prevents us from directly accessing 
the expected long-string phase near the Hagedorn 
energy density via AdS regulated equilibrium 
configurations. On the other hand, the AdS/CFT 
correspondence provides an indirect view of the
high-energy phase via the dual description of
large black holes in terms of a gauge theory in a 
deconfined phase.}
If there is a region outside a large AdS 
black hole that is characterized by a proper 
temperature of order the Hawking temperature
we would have a problem. This is because the 
minimum Hawking temperature of an AdS black hole 
is of order the AdS scale and in the limit of very large
black holes the Hawking temperature becomes 
arbitrarily high, higher than the Hagedorn temperature
even. Such high temperatures would precipitate black 
hole nucleation outside the horizon of the large black 
hole leading to an instability. 

We will argue that the above problems are avoided
and that an observer in free fall outside a large AdS
black hole does not see temperatures anywhere 
near the Hawking temperature.
In fact, to such an observer, a very large 
AdS--Schwarzschild black hole is an extremely cold 
object surrounded by almost empty space. We draw
this conclusion from an order-of-magnitude analysis 
of the semiclassical corrections to the classical black 
hole geometry. The corrections to the
total energy of the system include a contribution
from the radiation bath surrounding the black hole
and based on the size of these corrections we conclude
that the energy density of the radiation bath goes to
zero in the limit of infinite black hole mass. 

Our indirect calculation indicates that large 
AdS--Schwarzschild black holes have more in common 
with Schwarzschild black holes in asymptotically 
flat spacetime than the enormous
difference in Hawking temperatures would suggest. 
From the point of view of semiclassical gravity a large
AdS--Schwarzschild black hole appears to be a cold
object that emits very little Hawking radiation, just like
we know to be the case for an ordinary Schwarzschild 
black hole of the same mass in a background with
zero cosmological constant. The very high Hawking
temperature of a large AdS--Schwarzschild black hole
characterizes a dual thermal state in a gauge theory
via the AdS/CFT correspondence but on the gravity side
this temperature is not realized by observers in free fall
anywhere in the spacetime geometry outside a large 
black hole. The statement that large AdS--Schwarzschild 
black holes have positive specific heat is thus primarily 
a statement about the dual gauge theory. 

\section{Classical solutions}

In this section we present the classical solutions
that will enter into our subsequent discussion. We take 
spacetime to be four-dimensional. This is mostly to keep 
the notation simple but it will also be convenient later on
when we consider semiclassical corrections since most of
the existing literature on spherical reduction deals
with four-dimensional theories of gravity. It is 
straightforward to generalize the classical solutions to 
higher dimensions.

\subsection{AdS--Schwarzschild black holes}
\label{sec:classical}
The AdS--Schwarzschild metric in four dimensions reads 
\beq
\label{eq:metric}
ds^2 = -f_0 (r) dt^2 + \frac {dr^2}{f_0 (r)} + r^2 d\Omega^2_2 \,,
\eeq 
where the radial function $f_0 (r)$ is 
\beq
\label{classmetric}
f_0 (r) = \frac {r^2}{\ell^2} + 1 - \frac {\mu}{r} \,.
\eeq
For $\mu =0$ the metric reduces to that of four-dimensional
anti-de Sitter spacetime with radius of curvature $\ell$ and 
cosmological constant $\Lambda = - 3 \ell^{-2}$.
For $\mu > 0$ the metric describes an eternal black hole with 
an event horizon at $r=r_s$ where $r_s$ is the single real valued
zero of $f_0 (r)$. 
The parameter $\mu$ is proportional to the mass of the black 
hole and can be expressed in terms of the horizon radius $r_s$ 
as follows: 
\beq
\mu = r_s + \frac {r_s^3}{\ell^2} \,.
\eeq 
The Hawking temperature for an 
AdS$_4$--Schwarzschild black hole can be obtained by 
rotating to Euclidean signature in 
(\ref{eq:metric}) and finding the period of Euclidean time that
avoids a conical singularity at $r=r_s$. It is given by
\beq
T_{H} = \frac {1}{4\pi} \, f_0^\prime (r_s) 
= \frac {1}{4\pi} \left( \frac 1{r_s} + \frac {3r_s}{\ell^2} \right) \,.
\eeq
There is a minimum Hawking temperature which is of order 
the characteristic
energy scale of the AdS background, $T\geq \sqrt{3}/2\pi\ell$.
Each temperature value above the minimum one is
realized for two different values of $r_s$ so there are two
branches of AdS--Schwarzschild black holes: 'large' black
holes with $r_s > \ell/\sqrt{3}$, and 'small' black holes 
with $r_s < \ell/\sqrt{3}$.

Let us consider the length scales that enter into
our discussion. First of all there is the AdS length $\ell$ 
which we take to be large compared to any fundamental
length scale such as the string length, {\it i.e.} $\ell\gg\ell_s$.
This is in order to justify using classical geometry in the
first place and to keep quantum corrections under control
in our semiclassical considerations later on. 
It is also useful to have in mind an intermediate length scale 
$\ell_o$ that we associate with macroscopic observers who can
move around in spacetime, experience tidal effects, measure 
a local temperature, {\it etc.} We will be interested in 
static semi-classical solutions that describe large black
holes in equilibrium with an 'atmosphere' consisting of 
Hawking radiation or low-energy solutions that describe 
a non-singular 'star' made of self-gravitating radiation.
In either case, the geometry varies on length scales of 
order the AdS length $\ell$ or larger and conditions for 
local thermal equilibrium are satisfied at the $\ell_o$ scale
provided 
\beq
\ell_s \ll \ell_o \ll\ell \,.
\eeq

We will mostly be concerned with large 
AdS--Schwarzschild black holes and among those it
is interesting to consider the very large limit where
\beq
r_s \approx (\mu\ell^2)^{1/3} \gg \ell \,.
\eeq
Some features are universal in this limit. Consider for 
example the scalar invariant 
\beq
R_{abcd}R^{abcd}
= 12\left( \frac{2}{\ell^4}+ \frac{\mu^2}{r^6}\right) \,,
\eeq
where $R_{abcd}$ are components of the Riemann
tensor and the right hand side is the value obtained for the
AdS--Schwarzschild metric (\ref{eq:metric}). Now let 
$r=r_s$ and take the limit of a very large black hole to
obtain
\beq
R_{abcd}R^{abcd}\Big\vert_\textrm{\small horizon}
= \frac{36}{\ell^4}\left(1+ O(\ell /\mu)^{2/3}\right) \,,
\eeq
which is independent of the black hole mass in the
$\mu\gg\ell$ limit. In other
words the near-horizon region of very large black 
holes has the same AdS scale curvature for all (very
large) black hole masses. Note that this universal
curvature is not equal to the curvature of empty AdS
vacuum with the same cosmological constant but
rather an $O(1)$ multiple of the vacuum value. 

At the quantum level AdS black holes 
emit Hawking radiation \cite{Hawking:1974sw} 
but due to the confining gravitational 
potential of the asymptotic AdS background the Hawking
radiation cannot escape to infinity and is reflected back 
towards the black hole. The fate of an evaporating 
AdS--Schwarzschild black hole depends very much on
which of the two branches it belongs to. 
A large black hole reabsorbs most of the reflected 
Hawking radiation and eventually settles into an
equilibrium where the black hole 
continually exchanges radiation with a
surrounding 'atmosphere'. In Section~\ref{sec:semiclass}
we obtain static semiclassical solutions that
describe such equilibrium configurations.
A small AdS black hole, on the other hand, radiates away 
its mass in a runaway process like an ordinary 
Schwarzschild black hole in asymptotically flat spacetime. 
The Hawking radiation is reflected back from infinity but it 
is not reabsorbed by the black hole at a high enough rate
to keep up with the accelerating evaporation process. 
At the end of day the small black hole has completely 
evaporated and its energy is instead contained in a 
spherically symmetric 'star' made of self-gravitating 
radiation \cite{Page:1985em,Hubeny:2006yu}.

\subsection{Self-gravitating radiation}
\label{adsstars}
Classical solutions that describe a spherical concentration
of self-gravitating radiation in a spacetime with negative
cosmological constant are easily 
found~\cite{Page:1985em,Hubeny:2006yu}. 
Let us briefly review their 
construction in four spacetime dimensions.\footnote{The 
corresponding solutions in five-dimensional spacetime 
are discussed in detail in \cite{Hubeny:2006yu}.}  For 
simplicity we assume that the Hawking radiation consists 
of massless particles and model its gravitational effect by
a spherically symmetric perfect fluid energy-momentum tensor,
\beq
T_{ab} = \rho(r) u_a u_b +p(r)(g_{ab}+u_a u_b) \,,
\eeq
with a linear equation of state,
\beq
\rho(r) = 3 p(r) \,.
\eeq
Here $u_a$ is the 4-velocity of the radiation fluid.
This energy-momentum tensor is inserted into the
Einstein equations along with a spherically symmetric
metric ansatz,
\beq
ds^2 = - \frac{1}{\ell^2}
     \left(\frac{\rho_\infty}{ \rho (r)}\right)^{1/2} dt^2 
       + \frac{dr^2}{(\frac{r^2}{\ell^2} +1 - \frac{2 m(r)}{r})} 
       + r^2 d\Omega^2 \,.
\eeq
The normalization constant $\rho_\infty$, which is 
determined by the asymptotic large $r$ behavior of 
the energy density, 
\beq
\rho(r)  \approx  \frac{\rho_\infty}{r^4} \,,
\label{density}
\eeq 
is needed in order for the asymptotically 
AdS metric to have the right characteristic length $\ell$.
The Einstein equations then reduce to a pair of 
ordinary differential equations. One of the equations 
determines the so-called mass function in terms of the 
energy density,
\beq
\label{eq:massfcn}
\frac {dm(r)}{dr} = 4 \pi r^2 \rho(r) \,,
\eeq
and the other is the Oppenheimer--Volkoff equation,
\beq
\label{eq:ov}
\frac {d\rho(r)}{dr} = -
\frac{4\rho(r)}{r} \frac{ (\frac{4}{3}\pi r^3 \rho(r) + m(r) 
+ \ell^{-2} r^3)}{(\ell^{-2} r^3 + r - 2m(r) )}\,,
\eeq
with the effect of the negative cosmological constant 
included through the terms involving the AdS length 
scale $\ell$.

\EPSFIGURE{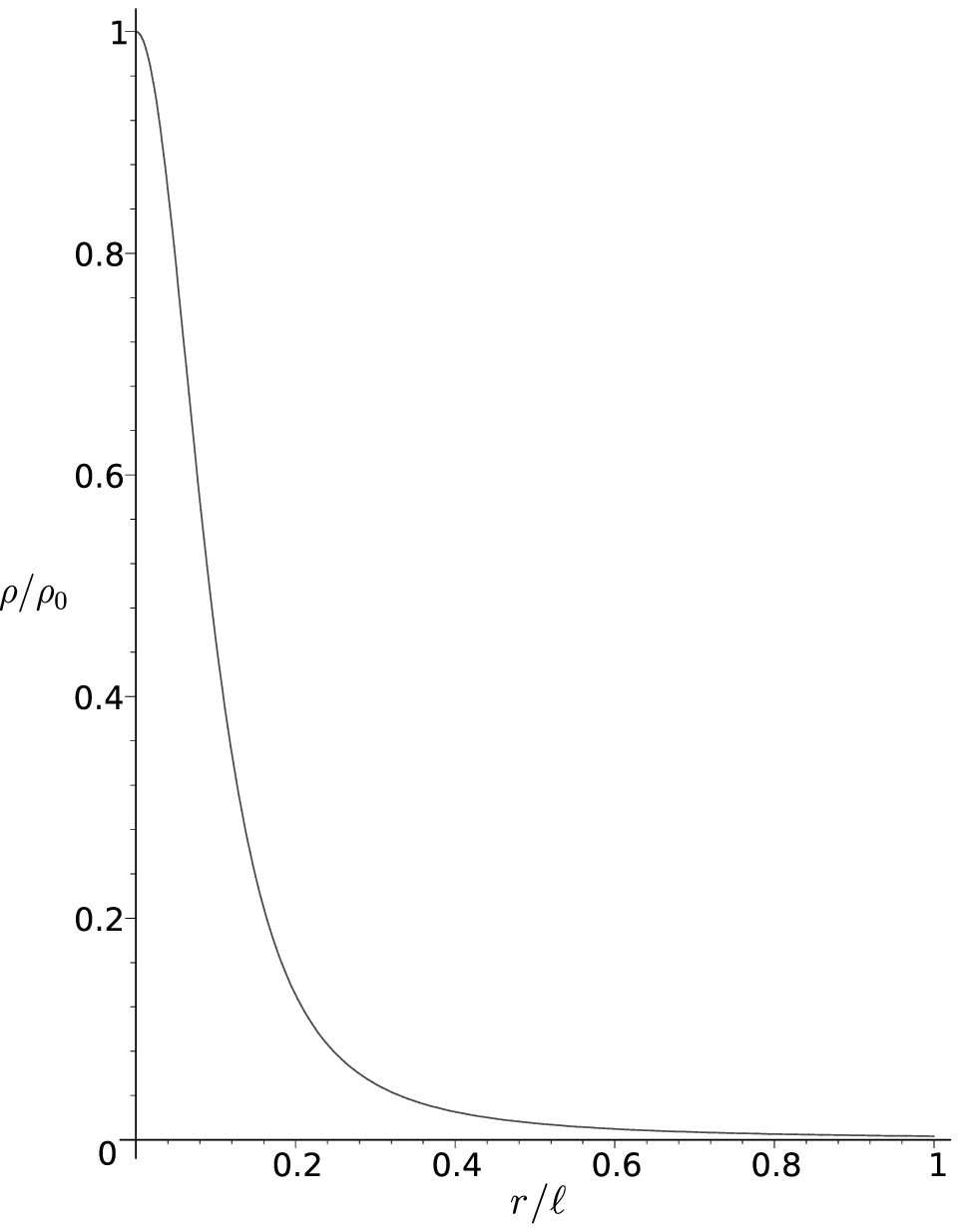, width=6.5cm}{
Density profile of spherically symmetric self-gravitating radiation.
\label{fig:one}}

The first order differential equations (\ref{eq:massfcn}) 
and (\ref{eq:ov}) can easily be solved numerically and 
turn out to have a one-parameter family of solutions
labeled by the central energy density $\rho(0)$. 
The solutions are non-singular and describe a 
spherically symmetric density profile that monotonically
decreases away from $r=0$, as shown in 
Figure~\ref{fig:one}.
The asymptotic behavior of the energy density for 
$r\gg\ell$ is of the form (\ref{density}), with $\rho_\infty$
determined by the value of $\rho(0)$.
The mass function vanishes at the origin, $m(0)=0$, 
and asymptotically it behaves as
\beq
m(\infty)-m(r)  =  O(r^{-1}) \,,
\eeq
where $m(\infty)$ denotes the total mass of the AdS 'star'.

Formally the central energy density has no upper 
bound but new physics comes into play when $\rho(0)$ 
becomes very large. In string theory, for example, the 
equation of state of the radiation fluid is modified 
to $p\approx 0$ at very high energy densities 
approaching the Hagedorn scale reflecting the tendency 
to form long strings that exert very little pressure compared 
to a gas of massless particles at the same energy density. 

The  AdS star
solutions have two interesting features that were pointed
out in \cite{Hubeny:2006yu}. First of all $m(\infty)$ is 
bounded from above by a relatively small mass which
is of order one in units of the AdS length $\ell$.
This means in particular that one of these self-gravitating 
radiation configurations can at most have a total mass
that equals that of a small AdS--Schwarzschild black hole
and never that of a large AdS--Schwarzschild black hole. 
This is one reason why we expect self-gravitating radiation
to be the endpoint of evaporation of small black holes
while large black holes must settle into equilibrium 
with a surrounding radiation atmosphere.

The second curious feature observed in \cite{Hubeny:2006yu} 
is that the total mass is not a 
monotonic function of the central energy density, which is
the parameter that labels the 'star' solutions. Instead 
$m(\infty)$ reaches its maximum value for a finite value of 
$\rho(0)$ and then drops somewhat lower as 
$\rho(0)\rightarrow\infty$, oscillating about a limiting value.
It was speculated in~\cite{Hubeny:2006yu} that this corresponds 
to the onset of a radial instability as further discussed 
in~\cite{Vaganov:2007at,Hammersley:2007rp}. 

The conclusion that $m(\infty)$ is bounded from above is
unchanged when we consider a gas made of strings rather 
than a gas of massless particles. In fact, the modified equation 
of state of the string gas is expected to precipitate the transition 
to a black hole at a somewhat {\it lower\/} total energy.
This is because the 
$p\approx 0$ equation of state of the long string phase is too 
soft to support a non-singular AdS star once the energy density in 
the center reaches the string scale. 

\section{Semiclassical geometry} 
\label{sec:semiclass}
In this section we consider quantum corrections to adS black 
hole geometries. We look for static semiclassical solutions that 
describe spherically symmetric adS black holes in equilibrium 
with a surrounding atmosphere consisting of matter radiation.
We use spherical reduction to make the semiclassical calculation
tractable. At the end of the day we find that the energy density
in radiation surrounding a very large black hole is small which
suggests that the back-reaction problem is indeed dominated 
by s-wave modes and that the spherical approximation is 
justified.\footnote{The reliability of the s-wave approximation to
four-dimensional gravity is discussed in \cite{Nojiri:1999br}.}

\subsection{One-loop action}
We start from the 3+1 dimensional Einstein--Hilbert action with a 
negative cosmological constant,
\beq
S = \frac{1}{16\pi G}\int d^4x\> \sqrt{-g^{(4)}}
\left(R^{(4)} +\frac{6}{\ell^2}\right) \,,
\eeq
and use a metric ansatz,
\beq
ds^2 = g_{ij}(x^0,x^1)dx^i dx^j + r(x^0,x^1)^2 d\Omega_2^2 \,,
\eeq
to perform a spherical reduction to the
1+1 dimensional classical action
\beq
S_0 = \frac 1{4 G} \int d^2x \sqrt{-g} 
\left( r^2 R-4r\nabla^2r - 2(\nabla r)^2 + 2 + 6 r^2 \ell^{-2} \right) \,.
\label{action}
\eeq
We also include the spherically reduced action of $N$ scalar 
fields that are minimally coupled in 3+1 dimensions,
\beq
S_m = -2\pi \int d^2x\> \sqrt{-g} \,r^2 \sum_{i=1}^N (\nabla f_i)^2 \,.
\eeq
For $N\gg 1$ the one-loop effective action describing the 
semiclassical back-reaction due to this matter sector 
is given by adding the following conformal anomaly term to the action 
\cite{Callan:1992rs,Elizalde:1993ew,Mukhanov:1994ax,Chiba:1997ex,Bousso:1997cg,
Nojiri:1997hx,Ichinose:1997gq,Kummer:1997jr,Dowker:1998bm,Nojiri:2000ja},
\beq
S_q = - \frac N{12} \int d^2x \sqrt{-g} \left(
\frac 12 R \Box^{-1} R - \frac {6 (\nabla r)^2}{r^2} \Box^{-1} R 
- 6 R \ln r \right) \,.
\eeq
We can express the back-reaction term in local form using 
auxiliary fields $\psi$ and $\chi$ \cite{BRM}: 
\beq
S_q = - \frac N{12} \int d^2x \sqrt{-g} 
\left( R(\psi - 6 \chi - 6 \ln r) + \frac 12 (\nabla \psi)^2 
       -6 \nabla \psi \nabla \chi - 6 \psi \frac {(\nabla r)^2}{r^2} 
\right)
\label{anom}
\eeq
Quantum corrections due to this term are small when 
$N G/\ell^2 \ll 1$. This can be arranged for any given values of 
$N\gg 1$ and $G$ by choosing a sufficiently large AdS length~$\ell$.  

\subsection{Static solutions}

We look for static solutions of the semiclassical equations of
motion in conformal gauge, $g_{\mu\nu} = f(x)
\eta_{\mu\nu}$ with the classical matter fields $f_i$ set to zero. 
The remaining fields only depend on the coordinate $x$
and the semiclassical equations reduce to 
\bea 
\psi^{\prime\prime} + ( \ln f )^{\prime\prime} &=& 0 \,, \label{psi_eq} \\
\chi^{\prime\prime} - \frac {(\rdot)^2}{r^2} &=& 0 \,, \label{chi_eq} \\
r^2 \left( \frac 12  (\ln f)^{\prime\prime} - \frac{3f}{\ell^2} + 
\frac {r^{\prime \prime}}r \right) &=& 
- N G \left( ( \psi \frac {\rdot}r )^\prime + \frac 12 (\ln f)^{\prime\prime} \right) , 
\label{f_eq} \\
(\rdot)^2 + r r^{\prime \prime} - f (1 + \frac{3r^2}{\ell^2}) &=& 
-N G \left( \frac 16 (\ln f )^{\prime\prime}+ \frac {r^{\prime \prime}}r \right) ,
\label{r_eq}
\eea 
where a prime denotes a derivative with respect to $x$.
The constraint equation corresponding to this gauge choice is
\bea
- r r^{\prime \prime} + r \rdot (\ln f)^\prime &=& 
N G \left( -\frac 16 \psi^{\prime\prime} +  \frac {r^{\prime \prime}}r 
+ (\ln f)^\prime (\frac 16 \psi^\prime -  \chip -  \frac {\rdot}r) \right. 
\nonumber \\ 
 & & \left. + \frac{1}{12} (\psi^\prime)^2 -  (\psi \chip)^\prime \right) .
\label{cs_eq}
\eea 

Our aim is to find the leading semiclassical correction to the mass of
a black hole. For this we need an approximate solution of these 
equations that is first order in $NG/\ell^2$ and valid for large $r$. 
Since there is no explicit
dependence on coordinate $x$ in the system (\ref{psi_eq})-(\ref{cs_eq}),
we look for a solution of the fields $\psi$, $\chip$, $f$ and $\rdot$ as
a function of the field $r(x)$. 

First of all it is straightforward to verify that the 
AdS--Schwarzschild spacetime,
\bea
\label{classol}
f(r) &=& f_0(r) = \frac {r^2}{\ell^2} + 1 - \frac {\mu}r \,,  \\
\rdot (r) &\equiv& \frac {dr}{dx} = f_0(r) \,, \label{rdotr}
\eea
is recovered as a solution of equations (\ref{f_eq})-(\ref{cs_eq}) 
when $N=0$.

Next we consider the back-reaction due to the one-loop correction term
using $NG/\ell^2$ as a small expansion parameter. The first step is to 
determine the $\OO(N^0)$ behavior of the auxilliary fields $\psi$ 
and $\chi$. Inserting the classical solution (\ref{classol}) for $f$ and 
$\frac{dr}{dx}$ into equations (\ref{psi_eq}) and (\ref{chi_eq}) leads to
\bea
\frac{d}{dr}(f_0 \frac{d\psi}{dr}+\frac{df_0}{dr}) &=& 0 \,, \label{psizeroeq}  \\
\frac{d}{dr} \chi^\prime  &=& \frac {f_0}{r^2} \,. \label{chizeroeq}
\eea
We do not need the explicit form of $\chi(r)$.  Only derivatives of $\chi$, 
and not the field itself, appear in the equations of motion and for our purposes 
it is enough to solve for $\chip$.

Equation (\ref{chizeroeq}) is easily integrated,
\beq
\label{chip_as}
\chip  = \frac {r}{\ell^2} + \chi_1 - \frac 1r + \frac {\mu}{2r^2} \,.
\eeq
To fix the constant of integration $\chi_1$ we  
require that $\chi$ and its derivatives with respect to $r$ be free of 
singularities at the horizon. From the relation
\beq
\chip \equiv\frac{d\chi}{dx} = f_0 \frac{d\chi}{dr} \,,
\eeq
which holds at the level of approximation that we are working with, 
we immediately see that $\chip$ must vanish at the horizon
$r=r_s$, {\it i.e.} that 
\beq
\chi_1 = \frac 12 \left( \frac 1{r_s} - \frac {3 r_s}{\ell^2} \right) .
\eeq
This expression diverges in the $r_s\rightarrow 0$ limit but 
this is not a problem because the semiclassical equations 
that we are working with are only valid for $r^2\gg NG$ and 
this puts a lower limit on the size of semiclassical black holes
that they apply to.

A first integral of equation (\ref{psizeroeq}) gives
\beq
f_0 \frac{d\psi}{dr}+\frac{df_0}{dr} = \psi_1 \,.
\eeq
The integration constant is fixed by the requirement that 
$d\psi/dr$ be non-singular at the horizon,
\beq 
\label{psi_smooth}
\psi_1 = \frac 1{r_s} + \frac {3 r_s}{\ell^2}  \,.
\eeq
A second integration of equation (\ref{psizeroeq}) then gives
\beq
\psi(r) = -\ln f_0(r) + \psi_1 \int^r \frac{d\tilde{r}}{f_0(\tilde{r})} \,.
\eeq
The asymptotic expansion of $\psi(r)$ at large $r$ is given by
\beq
\label{psi_as}
\psi(r) = -2\ln r + \psi_0 - \frac{\ell^2 \psi_1}{r} + \OO(r^{-2}) \,,
\eeq
where $\psi_0$ is another integration constant, which will
not affect any of our conclusions and can be left undetermined.

Next, we use the results (\ref{chip_as}) and (\ref{psi_as}) to 
solve the constraint equation (\ref{cs_eq}) to order
$\OO(NG/\ell^2)$. Writing a 
first-order ansatz for the fields $f(r)$ and $\frac{dr}{dx}$,
\beq
f(r) = f_0(r) + \frac {NG}{\ell^2} \, D(r) \>,\qquad 
\frac{dr}{dx} = f_0(r) + \frac {NG}{\ell^2} \, C(r) \>, 
\eeq
and expanding to order $\OO(NG/\ell^2)$, 
the constraint equation (\ref{cs_eq}) can be written 
\bea
\frac{d}{dr} \left( \frac {D(r)-C(r)}{f_0} \right) &=&  
\frac {\ell^2}{r f_0} \left( \frac{d^2f_0}{dr^2} + \frac{6}{r} \frac {df_0}{dr} 
+ (\frac{1}{f_0}\frac{df_0}{dr}) (\psi_1 - \frac{df_0}{dr} 
- 6 \chip - \frac {6}{r}f_0) \right. \nonumber \\ 
 & & \left. \quad
+ \frac 1{2 f_0} (\psi_1 - \frac{df_0}{dr})^2 - 6 \frac{d}{dr} ( \psi \chip )
\right) .
\eea 
Integrating this equation gives 
\beq
C(r) - D(r) = \ln r + \frac 12 - \frac 12 \psi_0 + \OO(r^{-2}\ln r) \,.
\label{ceedee}
\eeq
There is an integration constant which we have set to zero to
preserve the asymptotic value of the cosmological constant. 

Expanding equation (\ref{r_eq}) to order $\OO(NG/\ell^2)$, we 
obtain another linear equation involving $C(r)$ and $D(r)$,
\beq
r \frac{dC}{dr} + \left(2+\frac{r}{f_0}\frac{df_0}{dr}\right) C
- \left(1+\frac{3r^2}{\ell^2}\right) D = -\frac{\ell^2}{6}
\left(\frac{d^2f_0}{dr^2}+\frac{6}{r}\frac{df_0}{dr}\right) .
\eeq
By using the result  (\ref{ceedee}), we can finally solve for the
asymptotic expansions of $C(r)$ and $D(r)$,
\bea
C(r) &=& -3 \ln r - \frac 56 + \frac 32 \psi_0 + \frac {c_0}r + 
\OO(r^{-2}\ln r) \\ 
D(r) &=& -4 \ln r - \frac 43 + 2 \psi_0 + \frac {c_0}r + 
\OO(r^{-2}\ln r) 
\eea
where $c_0$ is an integration constant with dimensions of
length. We find it convenient to express it in terms of a
characteristic length scale of the black hole geometry,
\beq
c_0 = \alpha \, r_s \,,
\label{czero}
\eeq
with $\alpha$ an undetermined dimensionless constant.
It is straightforward to show that  with these asymptotic 
expansions for $C(r)$ and $D(r)$ the remaining equation 
of motion (\ref{f_eq}) is also satisfied to 
$\OO(NG/\ell^2)$, which provides a check on the solution
that we have found.

\section{Energy considerations}

In this section we  estimate the leading order quantum corrections 
to the total energy of a large AdS-Schwarzschild black hole and based 
on this we can estimate the energy density of the radiation atmosphere 
that surrounds the black hole. We define the total energy in terms
of a Brown-York quasilocal stress tensor~\cite{Brown:1992br} 
adapting the subtraction procedure of Balasubramanian and 
Kraus~\cite{BK} to our spherically reduced problem.\footnote{See
\cite{Aros:1999id,Olea:2005gb} for alternative approaches to regulating
conserved quantities in asymptotically AdS spacetimes.}
We first consider the classical 1+1 dimensional theory and confirm 
that the formalism gives the correct mass for a classical 
adS-Schwarzschild black hole. We then include $\OO(NG/\ell^2)$
corrections and obtain an order of magnitude estimate of the
total energy residing in the semiclassical black hole atmosphere.

\subsection{Classical total energy}
The 1+1 dimensional classical action (\ref{action}) needs to be
supplemented by a boundary term in order to have a well posed
variational problem for metric variations $\delta g_{\mu\nu}$ 
that vanish at the $r\rightarrow \infty$ boundary.
This can ultimately be traced to the 
$r^2\, R$ and $r\, \nabla^2 r$ terms in (\ref{action}), both of
which contain second derivatives acting on field variables.
The boundary term whose variation cancels the unwanted 
terms from the variation of the bulk classical action can be
written
\beq
S_b = \frac{1}{2G}\int dt \sqrt{-\gamma}
\left(r^2\gamma^{tt}\nabla_t n_t + n^\mu \nabla_\mu(r^2)\right) .
\label{bterm1}
\eeq
Here we are using the two-dimensional coordinate time $t$
as the parameter along the boundary and the one-dimensional
boundary metric $\gamma_{tt}$ is induced from the two-dimensional 
$g_{\mu\nu}$. The covariant derivatives in this expression are 
with respect to the two-dimensional metric and $n^\mu$ are the 
components of the unit normal to the boundary. The combination
$\gamma^{tt}\nabla_t n_t$ is the one-dimensional analog of
the trace of the extrinsic curvature of the boundary while the 
$n^\mu \nabla_\mu(r^2)$ term can be traced to the spatial 
curvature of the two-spheres that we are reducing on.
The boundary term (\ref{bterm1}) can be written in a more
compact way,
\beq
S_b = \frac{1}{2G}\int dt \sqrt{-\gamma} \,
\nabla_\mu(r^2\,n^\mu)\, ,
\label{bterm2}
\eeq
where we have used that the tensor $\nabla_\mu n_\nu$ 
is tangential to the boundary. 

In conformal gauge, $g_{\mu\nu} = f(x) \eta_{\mu\nu}$, 
we find the following "bare" quasilocal stress tensor
\beq
T^{tt} = \frac{2}{\sqrt{-\gamma}} \frac{\delta S}{\delta\gamma_{tt}}
= \frac{1}{2G} \gamma^{tt}n^a\nabla_a (r^2)
=- \frac{1}{G}\, \frac{r}{\sqrt{f_0(r)}} \,,
\eeq
where we have used the classical value (\ref{rdotr}) for
the derivative 
$\frac{dr}{dx}$. This leads to a bare total energy,
\beq
\varepsilon = - \frac{r\,f_0(r)}{G} = 
\frac{1}{G}\left( -\frac{r^3}{\ell^2}- r +\mu\right)  \,,
\eeq
that diverges in the $r\rightarrow\infty$ limit. Following 
\cite{BK} we introduce a boundary counterterm in the
classical action to cancel this divergence. In one dimension
such a counterterm can only be a function of the
induced metric $\gamma_{tt}$ and the scalar field $r$ and
we find that 
\beq
S_{ct} = -\frac{1}{G\ell} \int dt \sqrt{-\gamma}
\left(r^2 + \frac{\ell^2}{2}\right) ,
\eeq
achieves the desired cancellation. With the addition of
this counterterm we obtain a finite total energy,
\beq
\varepsilon = \lim_{r\rightarrow\infty} 
\left(- \frac{r\,f_0(r)}{G} +\frac{\sqrt{f_0(r)}}{G\ell}(r^2+\frac{\ell^2}{2})
\right)
=\frac{\mu}{2G} \,,
\eeq
which equals the classical black hole mass.

\subsection{Semi-classical energy}
We now want to include leading order semiclassical
corrections in these expressions in order to estimate
the contribution of the black hole atmosphere to the
total energy. In the classical boundary term
(\ref{bterm2}) the function that accompanies the 
normal vector $n^\mu$ inside the parenthesis is
the function that multiplies the Ricci scalar in the
bulk classical action (\ref{action}). The natural
semi-classical generalization of the boundary term
is therefore
\beq
S'_b = \frac{1}{2G}\int dt \sqrt{-\gamma} \,
\nabla_\mu(h\,n^\mu)\, ,
\label{oneloopbterm}
\eeq
where $h=r^2 -\frac{NG}{3}(\psi-6\chi-6\log r)$ is the 
function that multiplies $R$ in the one-loop corrected bulk 
action $S_0+S_q$ in (\ref{action}) and  (\ref{anom}).

The semiclassical quasilocal stress tensor can now be
computed and one finds new divergences in the bare 
total energy. These divergences are canceled if we 
include a one-loop correction in the boundary
counterterm,
\beq
S'_{ct} = -\frac{\ell}{G} \int dt \sqrt{-\gamma}
\left(\frac{r^2}{\ell^2} + \frac{1}{2} 
- \frac{NG}{2\ell^2}(\psi+\frac{13}{3}) \right) .
\eeq
The remaining finite total energy is given by
\beq
\varepsilon = \frac{1}{2G}\left(\mu 
+\frac{NG}{\ell^2}((1-\alpha)r_s-\frac{5\ell^2}{3r_s} ) \right),
\eeq
where $\alpha$ comes from the undetermined integration 
constant in (\ref{czero}). 
For large black holes, with $r_s\gg \ell$, this reduces to
\beq
\varepsilon \approx \frac{1}{2G}\left(\mu 
+(1-\alpha)(\frac{NG}{\ell^2} )r_s \right).
\label{totalenergy}
\eeq
The $\OO(\frac{NG}{\ell^2})$ term in the total energy 
(\ref{totalenergy}) provides an order-of-magnitude estimate 
of the energy carried by the radiation bath surrounding
the black hole. This energy increases linearly with 
$r_s$ while the area of the black hole is proportional to 
$r_s^2$. It follows that the average proper energy 
density in the radiation outside the black hole must go 
to zero in the $r_s\rightarrow \infty$ limit.\footnote{It is of 
course possible that the leading behavior 
of the $\OO(\frac{NG}{\ell^2})$ term in the total energy 
should be interpreted as coming from a semiclassical 
correction to the mass of the black hole itself rather than
as energy carried by surrounding radiation but in this 
case the energy density of the radiation bath around a 
large AdS black hole is even smaller than our (already
very small) estimate.} 

\section{Discussion}

Our semiclassical calculation supports the notion that, 
despite having a very high Hawking temperature, a large 
AdS black hole is a cold object in the sense that observers 
in free fall outside the black hole do not see high-temperature 
thermal radiation. The absence of 
high-energy thermal radiation can also be seen via
a more direct calculation~\cite{eblt} that utilizes the global 
embedding of the AdS-Schwarzschild geometry into a 
higher dimensional flat spacetime along the lines 
of~\cite{Deser:1998xb}.

Our results have implications for string thermodynamics 
with an AdS infrared regulator~\cite{Barbon:2001di}. 
At low total energy the stable configuration of such a 
string gas is an AdS star in local thermal equilibrium.
Now consider increasing the total energy in order to 
study ever higher energy densities at the center of the 
AdS star. This can only be carried so far because once 
the energy density in the center becomes of order the 
string scale the system is unstable to gravitational collapse. 
This occurs at a total energy of order the AdS scale and
for higher total energies the stable configuration is a large 
AdS black hole.
The Hawking temperature of such a black hole can be 
arbitrarily high if the black hole is large enough but 
from the point of view of observers in free fall it seems
that we are not probing high temperatures at all.
The dual gauge theory does provide a thermal description
in terms of a high-temperature plasma of deconfined
gluons at the Hawking temperature but this is a very
indirect and non-local description of the gravitational
physics.

\end{document}